# Electronic and magnetic phase diagram in K$_x$Fe$_{2-y}$Se$_2$ superconductors

Y. J. Yan[1], M. Zhang[1], A. F. Wang[1], J. J. Ying[1], Z. Y. Li[1], W. Qin[1], X. G. Luo[1],

J. Q. Li[2], Jiangping Hu[2,3] and X. H. Chen[1]

1. Hefei National Laboratory for Physical Sciences at Microscale and Department of Physics, University of Science and Technology of China, Hefei, Anhui 230026, China

2. Beijing National Laboratory for condensed Matter Physics, Institute of Physics, Chinese Academy of Sciences, Beijing 100190, China

3. Department of Phyiscs, Purdue University, West Lafayette, Indiana 47907, USA

The correlation and competition between antiferromagnetism and superconductivity are one of the most fundamental issues in all of high temperature superconductors. The superconductivity in high temperature cuprate superconductors arises from suppressing an antiferromagnetic (AFM) Mott insulator phase by doping[1] while that in iron-pnictide high temperature superconductors arises from AFM semimetals and can coexist with AFM orders[2-9]. This key difference marked in their phase diagrams has raised many intriguing debates about whether the two materials can be placed in the same category to understand the mechanism of superconductivity. Recently, superconductivity at 32 K has been reported in iron-chalcogenide superconductors A$_x$Fe$_{2-y}$Se$_2$ (A=K, Rb, and Cs)[10-12], which have the same structure as that of iron-pnictide AFe$_2$As$_2$ (A=Ba, Sr, Ca and K)[13-15]. Here, we report electronic and magnetic phase diagram of K$_x$Fe$_{2-y}$Se$_2$ system as a function of Fe valence. We find two AFM insulating phases and reveal that the superconducting phase is sandwiched between them, and give direct evidence that the superconductivity in A$_x$Fe$_{2-y}$Se$_2$ originates from the AFM insulating parent compounds. The two insulating phases are

**characterized by two distinct superstructures caused by Fe vacancy orders with modulation wave vectors of $q_1$=(1/5, 3/5, 0) and $q_2$=(1/4, 3/4, 0), respectively. These experimental results strongly indicate that iron-based superconductors and cuprates share a common origin and mechanism of superconductivity.**

Understanding whether antiferromagnetism is responsible for superconductivity in high temperature superconductors continues to be one of the most important unresolved problems in modern condensed matter physics. The parent compounds of both iron-pnictide and curpate high temperature superconductors display strong antiferromagnetism[1,5,6,16-18]. However, for iron-pnitides, the parent compounds are semimetal and the superconductivity can coexist with AFM order[2-9] while for curpates, the parent compounds are AFM Mott-insulators and the superconductivity is developed after the AFM order is completely suppressed by doping[1,16]. This critical difference has led to strong debate whether the iron-pnictides are weak coupling electron systems or similar strong correlated electron systems as cuprates. In the weak coupling approach, the AFM in iron-pnictides is generated by the nesting between the electron-like Fermi surfaces near the zone corner and the hole-like Fermi surfaces near the zone center in the Brillouin zone. It is also shown that the scattering between the hole and electron pockets drives an $s^{\pm}$ pairing symmetry, a s-wave pairing symmetry characterized by opposite signs between the hole and electron Fermi surface[19-25]. In the strong coupling approach for iron-pnictides, the AFM order is mainly contributed from local spin moments and a similar s-wave pairing symmetry is predicted[23,26], originated from the strong AFM exchange coupling between two next nearest-neighbor iron atoms.

The newly discovered iron chalcogenide superconductors $A_xFe_{2-y}Se_2$ conceptually challenge the weak coupling picture in iron-pnictides because only electron Fermi surfaces around the zone corners are observed so that the scattering between the hole and electron pockets can not be responsible for the superconductivity in these new materials[27,28]. Moreover, the new materials display many intriguing physical properties: (i) They have strong insulating phase featured by eight orders of magnitude increase of resistance from high temperature to low temperature as shown in Fig.4(a) and Fig.4(b); (ii) In contrast to the iron pnictides, they develop an antiferromagnetism with Neel temperature ($T_N$) as high as 559 K and the ordered magnetic moment of more than $3\mu_B$[29]; (iii) they also display iron vacancy order at a temperature Ts=578 K, higher than $T_N$[29]. The existence of strong insulating phase and the large magnetic moment reflect that the physics in $A_xFe_{2-y}Se_2$ is rather local. However, we still lack of a systematic phase diagram of $K_xFe_{2-y}Se_2$ system to directly reveal the correlation and competition between AFM and superconductivity and reflect possible local strong correlation physics. Difficulty to obtain the phase diagram is mainly due to the existence of intrinsic Fe vacancies. The presence of Fe vacancy order[30] causes non-stoichiometry of $A_xFe_{2-y}Se_2$, so that the study on superconductivity and the determination of their parent compound become more complicated and difficult.

Here, we report the electronic and magnetic phase diagram of $K_xFe_{2-y}Se_2$ system as a function of Fe valence state. The Fe valence state is systematically controlled by changing the x and y in $K_xFe_{2-y}Se_2$ system. We found that phase diagram consists of three regions with distinct physical properties by measuring the transport, magnetic and superconducting properties and Fe vacancy order in the three regions of phase diagram using resistivity, Seebeck coefficient, magnetic susceptibility, transmission electron microscopy (TEM) measurements. We indentify that there exist two insulating phases with a gap of larger than 0.3 eV, and the superconductivity occurs in a narrow region of Fe valence ($V_{Fe}$) from 2 to ~1.94 between the two insulating phases.

Our main results are shown in Fig.1a, which maps out the detailed electronic and magnetic phase diagram against the average valence of iron for $K_xFe_{2-y}Se_2$ system. The average valence of iron from 1.86 to 2.06 is obtained from tens of single crystals with nominal different compositions of $A_xFe_{2-y}Se_2$ by changing the x and y (see Table I). First of all, all the samples show a similar behavior of paramagnetic weak metal above a certain temperature ($T_S$), at which a structural transition takes place due to the formation of the Fe vacancy order. Just slightly below $T_S$, an antiferromagnetic transition happens at a temperature $T_N$. Below $T_N$, there are three regions with respect to the average valence of iron for all the samples. In the region I with $V_{Fe} \geq 2.00$, an insulating state with long range AFM order is observed. In the region II with $1.935 < V_{Fe} < 2.00$, a superconducting state is observed together with a long range AFM order. The superconducting transition temperature ($T_C$), around 30 K, is robust against the Fe valence. In the region III with $V_{Fe} < 1.935$, another insulating state with a long range AFM order is formed. In the region I and II, the Ts and $T_N$ are very robust against the Fe valence, while decrease obviously with reducing $V_{Fe}$ in the region III.

We further confirm the phase diagram by measuring the Seebeck coefficients. The Seebeck coefficient at 300K is plotted against the Fe valence in Fig.1b. The typical Seebeck coefficient for the samples in the three regions is shown in Fig.2. The coefficients are large positive and large negative values in the two insulating phases $V_{Fe} > 2.00$ and $V_{Fe} < 1.935$, respectively. The result indicates the opposite types of the dominant charge carriers in the two insulating phases. In the region II, very small values of Seebeck coefficients were observed at 300 K. A divergent behavior in Seebeck coefficient occurs in the boundaries between region I and II, and region III and II. Such evolution of Seebeck coefficients with $V_{Fe}$ suggests the existence of Lifshitz transitions as the system moves into the superconducting region from the two insulating sides with the sudden change of the Fermi surface at the boundaries.

To understand the difference between two insulating phases, we use TEM to examine iron vacancy order in the region I and III. Figure 3 shows the typical TEM observations on the insulating crystals in the two regions. Fig. 3a shows a high-resolution TEM image taken from thin crystal in the region I, in which the ordered behavior as visible periodic features within the *a-b* plane can be clearly read out. Superstructure spots with the main diffraction spots are clearly illustrated, and are believed to originate from the Fe vacancy order. Careful examination reveals that the satellite spots in general are clearly visible in the *a\*-b\** plane of reciprocal space as clearly illustrated in Figs. 3b and can be characterized by a unique modulation wave vector $q_1 = (1/5, 3/5, 0)$ for the insulating samples of the region I. For the samples in region III, typical electron diffraction pattern taken along the [001] zone-axis directions exhibits the superstructure reflections within in the *a\*-b\** plane with modulation wave vector $q_2=(1/4, 3/4, 0)$, as shown in Fig.3c. The small arrow indicates (1/2, 1/2, 0) spot which may be due to K order. These results definitely show distinct properties in the two insulating phases.

In the region II of phase diagram, both superconductivity and antiferromagnetism exist below 30 K. Although our experimental results can not absolutely determine whether the two orders microscopically coexist or exist in the manner of phase separation, the latter is most likely the case. The current evidence supporting the microscopic coexistence is from Muon-spin rotation/relaxation (μSR) experiments in $Cs_{0.8}(FeSe_{0.98})_2$ crystal[31] and the measurements of magnetization and resistivity in single crystals $A_{0.8}Fe_{2-y}Se_2$ (A= K, Rb, Cs, Tl/K and Tl/Rb)[32]. However, a phase separation is evidenced by the intergrowth of single crystal with different c-axis lattice parameter reported by Luo et al[33]. Such intergrowth leads to two sets of reflections along

[001] direction as shown in the inset of Fig. 4(d). In addition, TEM observation has revealed the structural inhomogeneity of the Fe vacancy order. The superconductivity takes place in some areas with a basic tetragonal lattice without a long rang Fe vacancy order while the Fe vacancy order with modulation wave vector $q_1$=(1/5, 3/5, 0) or wave vector $q_2$=(1/4, 3/4, 0) can be observed in other domains[34]. These results suggest a microscopic phase separation in superconducting samples.

Finally, we discuss which insulating phase can be considered as the parent compound of $A_xFe_{2-y}Se_2$ superconductors. We study the physical properties for the samples with nominal compositions $A_{1-x}Fe_{1.5+x/2}Se_2$ (x=0, 0.1, 0.2, and 0.3). This series of samples with $A_{1-x}Fe_{1.5+x/2}Se_2$ keep the nominal valence of Fe to be +2. Figure 4a shows the temperature dependence of the resistivity for the samples of $A_{1-x}Fe_{1.5+x/2}Se_2$. Surprisingly, all these samples show insulating behavior with activation energy of about 100 meV. Resistivity exhibits a rapid increase around 550 K, and subsequent insulating behavior to the low temperature. The resistivity increases by about eight orders of magnitude with decreasing temperature from 500 K to 60 K. No superconductivity can be observed for this series of samples. These results suggest that the samples of $A_{1-x}Fe_{1.5+x/2}Se_2$ with nominal Fe valence of 2 are potential parent compound for the superconductors $A_xFe_{2-y}Se_2$. To further determine the parent compound, we focus on the two series of samples with nominal composition $A_xFe_ySe_2$ (x=1 and 0.8). Figure 4b shows the resistivity as a function of temperature in $KFe_ySe_2$ (y ranging from 1.5 to 2.4). All the samples show insulating behavior and no trace for superconductivity can be found. A sharp rapid increase in resistivity takes place in the temperature ranging from 525 K to 550 K due to the Fe vacancy order and AFM transition. The elemental analysis (as shown in Table I) indicates that the actual K concentration for all the samples is nearly 1, and Fe content is around 1.6, so that the average Fe valence always keep less than 1.94. These samples fall in the region III of phase diagram. This is why we cannot observe superconductivity for the samples KFe$_y$Se$_2$ with various Fe contents. It suggests that K content is very important for superconductivity. Now we take the insulating K$_{0.8}$Fe$_{1.6}$Se$_2$ as parent compound and try to dope it to induce superconductivity by adding more Fe in the starting material for crystal growth. Figure 4c shows the in-plane resistivity for the crystals with the nominal compositions K$_{0.8}$Fe$_{1.6+y}$Se$_2$ (y =0.1-1.4). All the samples show similar resistivity behavior. At the temperature around 550 K, resistivity shows a rapid increase and subsequently semiconductor-like behavior with decreasing temperature. With further decreasing temperature, resistivity exhibits metallic behavior, and the superconductivity shows up around 30 K although the samples still show AFM transition above 500 K (see supplementary information) as well as quite large value of resistivity in the normal state. A fully superconducting shielding volume fraction is observed by susceptibility for the samples of K$_{0.8}$Fe$_{1.6+y}$Se$_2$ as shown in Fig.4d. Element analysis indicates that the samples with nominal composition K$_{0.8}$Fe$_{1.6+y}$Se$_2$ falls in the region I of phase diagram for y<0.1, while in the region II of phase diagram for y≥0.1. These results prove that K$_{0.8}$Fe$_{1.6}$Se$_2$ is the parent compound of A$_x$Fe$_{2-y}$Se$_2$ superconductors.

In summary, we determine the electronic and magnetic phase diagram of A$_x$Fe$_{2-y}$Se$_2$ and show that the superconductivity originates from insulating phase of the region I in the phase diagram, similar to curpate superconductors. Superconductivity develops with doping electrons into the insulating phase of the region I. With further doping electron into system, another insulating phase of the region III shows up. The insulating phase in the region III could arise from the Fe vacancy order. Our findings cast new insight on the origin and mechanism of superconductivity and build a new bridge between cuprates and iron-based high temperature superconductors.

**Methods:**

$K_xFe_ySe_2$ single crystals used in this study were grown by using Bridgman method[11]. The compositions of crystals were determined using an energy-dispersive X-ray spectrometer (EDS) mounted on the field emission scanning electronic microscope (FESEM), Sirion200. At least five spots for each crystal have been measured to obtain the average potassium and iron concentration by considering selenium as 2. To make sure that the obtained compositions from EDS are consistent with each batch, more than two pieces of crystal for each sample from the same batch were used to determine the composition. The average valence of iron was obtained for each crystal by calculation with formula: (4-x)/y assuming without Se vacancy, where x and y were the actual concentration of potassium and iron from elemental analysis. All these results have been listed in Table I. We measured resistivity using standard four-probe method. For resistivity below 400 K, the measurements were carried out by using *Quantum Design* PPMS-9. The measurement of high temperature resistivity above 300 K was performed by using LR700 alternative current Resistance bridge with Type-K Chromel-Alumel thermocouples as thermometer in a home-built high-temperature oven. Magnetic susceptibility was measured by using *Quantum Design* MVSM-MPMS. High-temperature magnetic susceptibility was measured using high-temperature oven in a *Quantum Design* SQUID-MPMS-7. The Seebeck coefficients were measured on *Quantum Design* PPMS-9 with steady-state method by means of heat off and on mode. Specimens for TEM observation were prepared by peeling off a very thin sheet of a thickness around several tens microns from the single crystal and then milling by Ar ion. Microstructure analyses were performed on a FEI Tecnai-F20 TEM equipped with double-tilt cooling holder.

**Acknowledgements:** X.H.C would like to thank Profs. Z. Y. Weng and D. L. Feng for helpful discussion. This work was supported by the Natural Science Foundation of China, and by the Ministry of Science and Technology of China and Chinese Academy of Sciences.

**Author information:** Correspondence and requests for materials should be addressed to X.H.C. (chenxh@ustc.edu.cn).

**Table I |** The nominal and actual compositions, onset temperature of superconducting transition ($T_C$), AFM transition temperature ($T_N$), structure transition temperature ($T_S$) and Seebeck coefficient at 300 K ($S_{300\,K}$) for the $K_xFe_{2-y}Se_2$ single crystals.

| Nominal composition | Actual composition | Valence of iron | $T_c$ (K) | $T_N$ (K) | $T_S$ (K) | $S_{300K}$ (µV/K) |
|---|---|---|---|---|---|---|
| $KFe_{1.5}Se_2$ | $K_{0.99}Fe_{1.61}Se_2$ | 1.869 | 0 | 524 | 537 | -146 |
| $KFe_{1.55}Se_2$ | $K_{0.98}Fe_{1.62}Se_2$ | 1.864 | 0 | 519 | 530 | -167 |
| $KFe_2Se_2$ | $K_{0.98}Fe_{1.61}Se_2$ | 1.876 | 0 | 530 | 539 | -182 |
| $KFe_{2.4}Se_2$ | $K_{0.96}Fe_{1.60}Se_2$ | 1.900 | 0 | 534 | 544 | -212 |
| $K_{0.9}Fe_{1.5}Se_2$ | $K_{0.91}Fe_{1.60}Se_2$ | 1.931 | 0 | 545 | 554 | -153 |
| $K_{0.9}Fe_{1.55}Se_2$ | $K_{0.89}Fe_{1.62}Se_2$ | 1.920 | 0 | 534 | 545 | -187 |
| $K_{0.9}Fe_{1.7}Se_2$ | $K_{0.89}Fe_{1.61}Se_2$ | 1.932 | 0 | 537 | 550 | -360 |
| $K_{0.8}Fe_{1.7}Se_2$ | $K_{0.76}Fe_{1.65}Se_2$ | 1.964 | 31.4 | 541 | 549 | -56 |
| $K_{0.8}Fe_{1.75}Se_2$ | $K_{0.73}Fe_{1.68}Se_2$ | 1.946 | 30.5 | 538 | 550 | -48 |
| $K_{0.8}Fe_{1.8}Se_2$ | $K_{0.71}Fe_{1.65}Se_2$ | 1.994 | 30.1 | 539 | 547 | -7.7 |
| $K_{0.8}Fe_2Se_2$ | $K_{0.73}Fe_{1.67}Se_2$ | 1.958 | 31.8 | 534 | 546 | -4.0 |
| $K_{0.8}Fe_{2.2}Se_2$ | $K_{0.76}Fe_{1.67}Se_2$ | 1.940 | 31.8 | 538 | 548 | -0.5 |
| $K_{0.8}Fe_{2.4}Se_2$ | $K_{0.73}Fe_{1.65}Se_2$ | 1.982 | 31.4 | 536 | 544 | 3.4 |
| $K_{0.8}Fe_{2.6}Se_2$ | $K_{0.76}Fe_{1.63}Se_2$ | 1.988 | 32.8 | 540 | 547 | 7.0 |
| $K_{0.8}Fe_3Se_2$ | $K_{0.73}Fe_{1.66}Se_2$ | 1.970 | 32.6 | 535 | 545 | 5.7 |
| $K_{0.8}Fe_{1.6}Se_2$ | $K_{0.79}Fe_{1.60}Se_2$ | 2.006 | 0 | 543 | 551 | 175 |
| $K_{0.75}Fe_{1.6}Se_2$ | $K_{0.75}Fe_{1.60}Se_2$ | 2.031 | 0 | 538 | 548 | 135 |
| $K_{0.7}Fe_{1.65}Se_2$ | $K_{0.69}Fe_{1.61}Se_2$ | 2.056 | 0 | 546 | 555 | 113 |

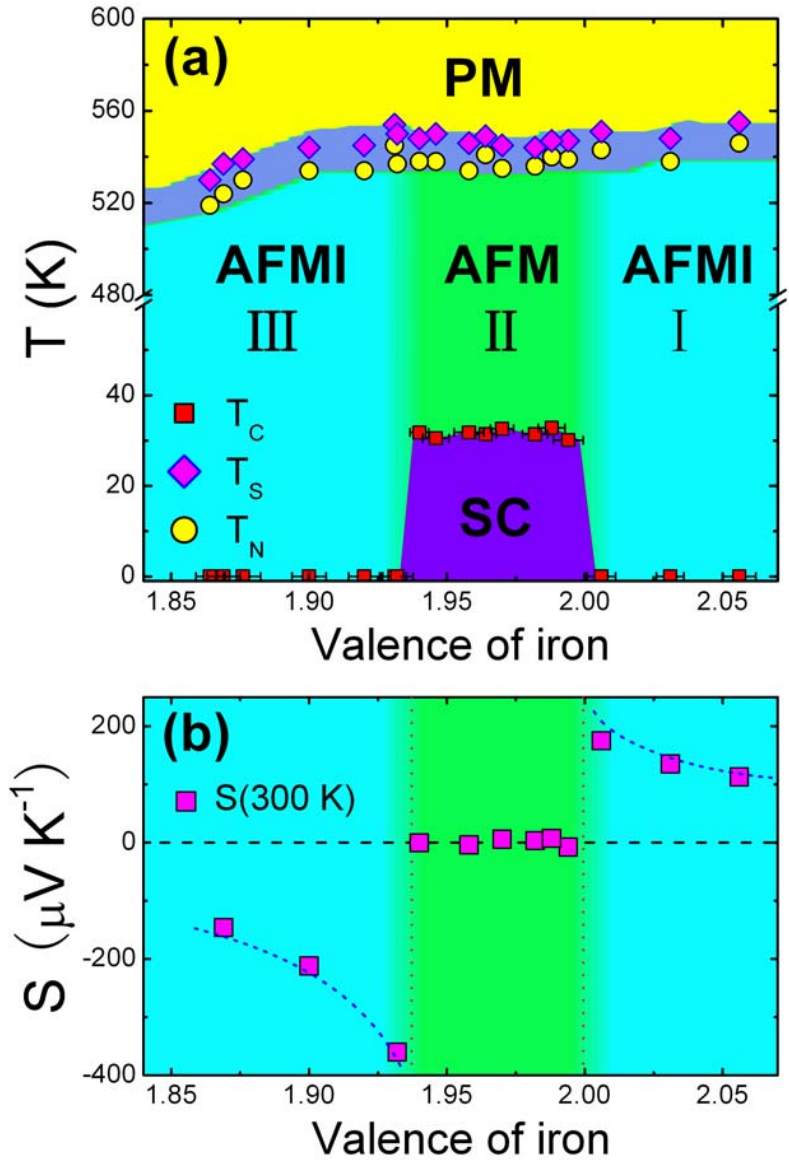

**Figure 1 | Electronic and magnetic phase diagram of $K_xFe_{2-y}Se_2$ as a function of Fe valence.**
(a): The phase diagram plotted against the valence of iron. The Néel temperature ($T_N$) of the AFM transition determined by magnetic susceptibility (yellow circles), the superconducting transition ($T_C$) obtained by resistivity and susceptibility (red squares), the temperature of the structural transition due to Fe vacancy ordering determined from the derivative of resistivity; PM: paramagnetic metal; SC: superconducting state; AFMI: antiferromagnetic insulator; (b): Evolution of Seebeck coefficients at 300 K (magenta squares) with the valence of iron.

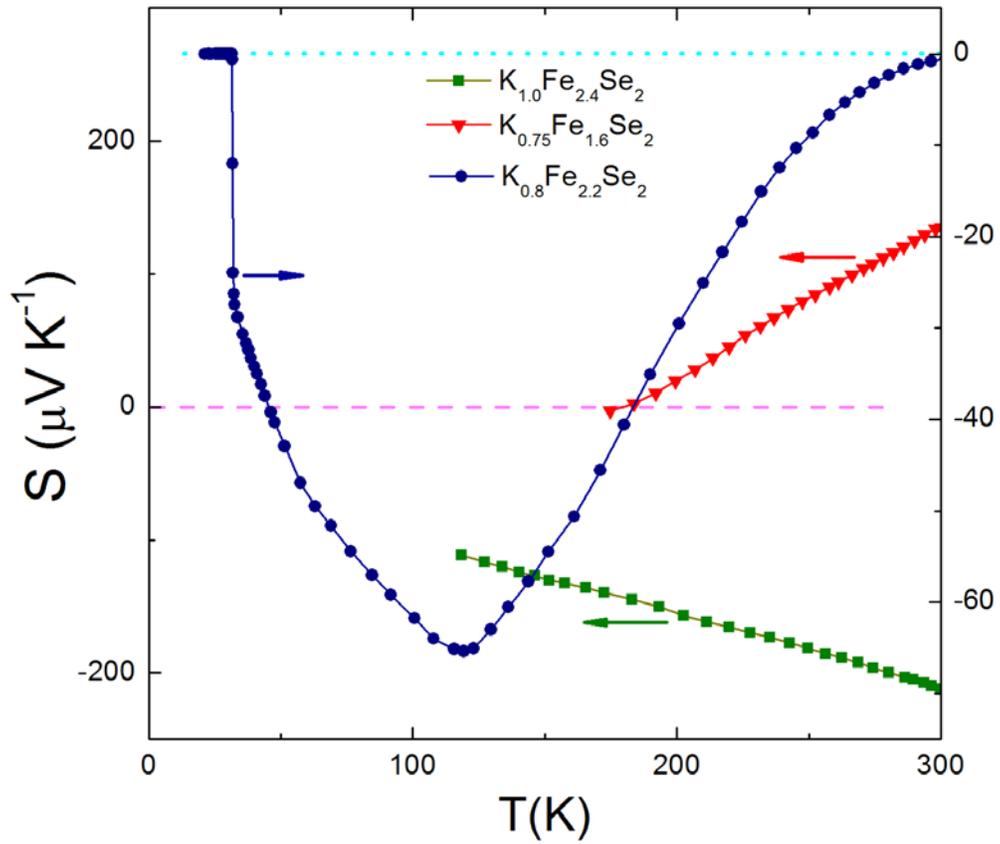

**Figure 2 | Typical Seebeck coefficient as a function of temperature for the samples in the three regions of the phase diagram.** Region I: $K_{0.75}Fe_{1.6}Se_2$; Region II: $K_{0.8}Fe_{2.2}Se_2$; Region III: $K_{1.0}Fe_{2.4}Se_2$.

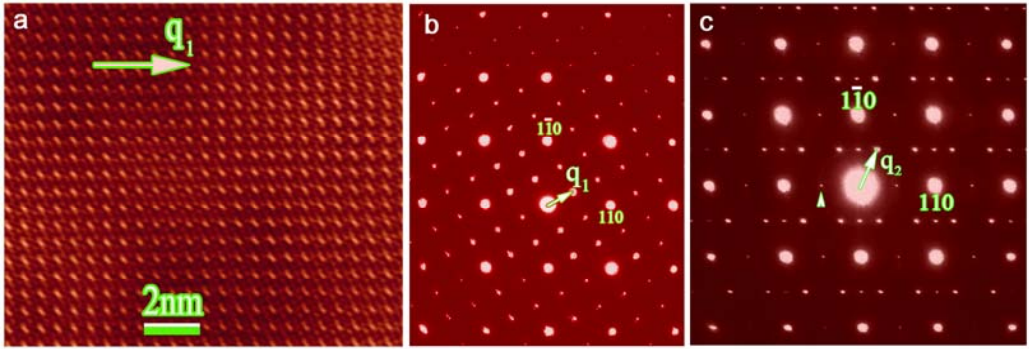

**Figure 3** |**Typical high-resolution TEM image and electron diffraction patterns for the samples in the two insulating phases. (a):** High-resolution TEM image of $K_{0.8}Fe_{1.6}Se_2$ in the region I taken along the [001] zone-axis direction, in which the ordered behavior as visible periodic features within the ***a-b*** plane can be clearly read out; **(b):** Electron diffraction pattern of $K_{0.8}Fe_{1.6}Se_2$ in the region I taken along the [001] zone-axis direction, superstructure spots are clearly visible in the ***a\*-b\**** plane of reciprocal space and can be characterized by a unique modulation wave vector $q_1$ = (1/5, 3/5, 0); **(c):** Electron diffraction pattern of $KFe_{1.5}Se_2$ in the region III taken along the [001] zone-axis direction, showing the superstructure reflections within the *a-b* reciprocal plane with wave vector $q_2$=(1/4, 3/4, 0),  and small arrow indicates (1/2, 1/2, 0) spots.

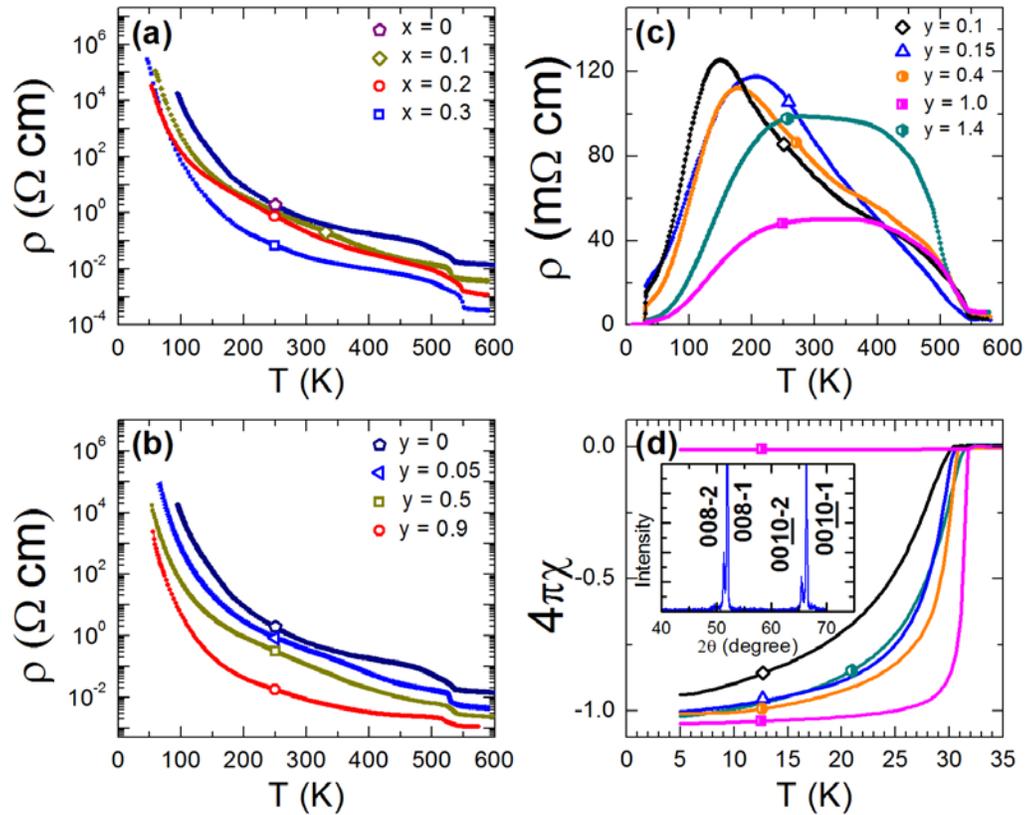

**Figure 4 | Evolution of transport properties and superconductivity from region I to region III in the phase diagram. (a):** Temperature dependence of the in-plane resistivity for the insulating samples with the nominal composition of $K_{1-x}Fe_{1.5+x/2}Se_2$ (x=0, 0.1, 0.2, 0.3); **(b):** In-plane resistivity as a function of temperature for the insulating crystals with nominal composition 0f $KFe_{1.5+y}Se_2$ (y = 0, 0.05, 0.5, 0.9) in the region III; **(c):** Temperature dependence of the resistivity for the crystals grown with the nominal composition of $K_{0.8}Fe_{1.6+y}Se_2$ (0.1 ≤ y ≤ 1.4) in region II. **(d):** The magnetic susceptibility data measured with a zero-field-cooling (ZFC) process and a field of 10 Oe applied within the *ab*-plane for the same crystals as that measured in (c). The field-cooling (FC) susceptibility with magnetic field of 10 Oe applied with the *ab*-plane fails to show Meissner effect for the $K_{0.8}Fe_{2.6}Se_2$. The inset of **d** shows the reflections of single crystal XRD taken at 300 K for $K_{0.8}Fe_2Se_2$ crystal, and two sets of reflections are observed, suggesting an intergrowth behavior and two phases in superconducting crystal.